\documentclass[twocolumn,showpacs,preprintnumbers]{revtex4}
\usepackage{amsmath}
\usepackage{amssymb}
\usepackage{amsfonts}
\usepackage{graphicx}
\usepackage{dcolumn}
\usepackage{bm}

\input{tcilatex}

\begin{document}

\title{Quantum Econophysics}
\author{Esteban Guevara Hidalgo$^{\dag \ddag }$}
\affiliation{$^{\dag }$Departamento de F\'{\i}sica, Escuela Polit\'{e}cnica Nacional,
Quito, Ecuador\\
$^{\ddag }$SI\'{O}N, Autopista General Rumi\~{n}ahui, Urbanizaci\'{o}n Ed%
\'{e}n del Valle, Sector 5, Calle 1 y Calle A \# 79, Quito, Ecuador}

\begin{abstract}
The relationships between game theory and quantum mechanics let us propose
certain quantization relationships through which we could describe and
understand not only quantum but also classical, evolutionary and the
biological systems that were described before through the replicator
dynamics. Quantum mechanics could be used to explain more correctly
biological and economical processes and even it could encloses theories like
games and evolutionary dynamics. This could make quantum mechanics a more
general theory that we had thought.

Although both systems analyzed are described through two apparently
different theories (quantum mechanics and game theory) it is shown that both
systems are analogous and thus exactly equivalents. So, we can take some
concepts and definitions from quantum mechanics and physics for the best
understanding of the behavior of economics and biology. Also, we could maybe
understand nature like a game in where its players compete for a common
welfare and the equilibrium of the system that they are members.
\end{abstract}

\maketitle
\email{esteban\_guevarah@yahoo.es}

\section{Introduction}

Could it have an relationship between quantum mechanics and game theories?
An actual relationship between these theories that describe two apparently
different systems would let us explain biological and economical processes
through quantum mechanics, quantum information theory and statistical
physics.

\bigskip

We also could try to find a method which let us make quantum a classical
system in order to analyze it from a absolutely different perspective and
under a physical equilibrium principle which would have to be exactly
equivalent to the defined classically in economics or biology. Physics tries
to describe approximately nature which is the most perfect system. The
equilibrium notion in a physical system is the central cause for this
perfection. We could make use of this physical equilibrium to its
application in conflictive systems like economics.

\bigskip

The present work analyze the relationships between quantum mechanics and
game theory and proposes through certain quantization relationships a
quantum understanding of classical systems.

\section{The von Neumann Equation \& the Statistical Mixture of States}

An ensemble is a collection of identically prepared physical systems. When
each member of the ensemble is characterized by the same state vector $%
\left\vert \Psi (t)\right\rangle $ it is called pure ensemble. If each
member has a probability $p_{i}$ of being in the state $\left\vert \Psi
_{i}(t)\right\rangle $ we have a mixed ensemble. Each member of a mixed
ensemble\ is a pure state and its evolution is given by Schr\"{o}dinger
equation. To describe correctly a statistical mixture of states it is
necessary the introduction of the density operator%
\begin{equation}
\rho (t)=\sum_{i=1}^{n}p_{i}\left\vert \Psi _{i}(t)\right\rangle
\left\langle \Psi _{i}(t)\right\vert  \label{1}
\end{equation}%
which contains all the physically significant information we can obtain
about the ensemble in question. Any two ensembles that produce the same
density operator are physically indistinguishable. The density operator can
be represented in matrix form. A pure state is specified by $p_{i}=1$ for
some $\left\vert \Psi _{i}(t)\right\rangle ,i=1,...,n$ and the matrix which
represents it has all its elements equal to zero except one $1$ on the
diagonal. The diagonal elements $\rho _{nn}$ of \ the density operator $\rho
(t)$ represents the average probability of finding the system in the state $%
\left\vert n\right\rangle $ and its sum is equal to $1$. The non-diagonal
elements $\rho _{np}$ expresses the interference effects between the states $%
\left\vert n\right\rangle $ and $\left\vert p\right\rangle $ which can
appear when the state $\left\vert \Psi _{i}\right\rangle $ is a coherent
linear superposition of these states. Suppose we make a measurement on a
mixed ensemble of some observable $A$. The ensemble average of $A$ is
defined by the average of the expected values measured in each member of the
ensemble described by $\left\vert \Psi _{i}(t)\right\rangle $\ and with
probability $p_{i}$, it means $\left\langle A\right\rangle _{\rho
}=p_{1}\left\langle A\right\rangle _{1}+p_{2}\left\langle A\right\rangle
_{2}+...+p_{n}\left\langle A\right\rangle _{n}$ and can be calculated by
using%
\begin{equation}
\left\langle A\right\rangle =Tr\left\{ \rho (t)A\right\} \text{.}  \label{2}
\end{equation}%
The time evolution of the density operator is given by the von Neumann
equation%
\begin{equation}
i\hbar \frac{d\rho }{dt}=\left[ \hat{H},\rho \right]  \label{3}
\end{equation}%
which is only a generalization of the Schr\"{o}dinger equation and the
quantum analogue of Liouville's theorem.

\section{The Replicator Dynamics \& EGT}

Game theory \cite{1,2,3} is the study of decision making of competing agents
in some conflict situation. It has been applied to solve many problems in
economics, social sciences, biology and engineering. The central equilibrium
concept in game theory is the Nash Equilibrium which is expressed through
the following condition%
\begin{equation}
E(p,p)\geq E(r,p)\text{.}  \label{4}
\end{equation}%
Players are in equilibrium if a change in strategies by any one of them $%
(p\rightarrow r)$ would lead that player to earn less than if he remained
with his current strategy $(p)$.

\bigskip 

Evolutionary game theory \cite{4,5,6} has been applied to the solution of
games from a different perspective. Through the replicator dynamics it is
possible to solve not only evolutionary but also classical games. That is
why EGT has been considered like a generalization of classical game theory.
Evolutionary game theory does not rely on rational assumptions but on the
idea that the Darwinian process of natural selection \cite{7} drives
organisms towards the optimization of reproductive success \cite{8}. Instead
of working out the optimal strategy, the different phenotypes in a
population are associated with the basic strategies that are shaped by trial
and error by a process of natural selection or learning.

\bigskip 

The model used in EGT is the following: Each agent in a n-player game where
the $i^{th}$ player has as strategy space $S_{i}$ is modelled by a
population of players which have to be partitioned into groups. Individuals
in the same group would all play the same strategy. Randomly we make play
the members of the subpopulations against each other. The subpopulations
that perform the best will grow and those that do not will shrink and
eventually will vanish. The process of natural selection assures survival of
the best players at the expense of the others. The natural selection process
that determines how populations playing specific strategies evolve is known
as the replicator dynamics \cite{5,6,9,10}%
\begin{gather}
\frac{dx_{i}}{dt}=\left[ f_{i}(x)-\left\langle f(x)\right\rangle \right]
x_{i}\text{,}  \label{5} \\
\frac{dx_{i}}{dt}=\left[ \sum_{j=1}^{n}a_{ij}x_{j}-%
\sum_{k,l=1}^{n}a_{kl}x_{k}x_{l}\right] x_{i}\text{.}  \label{6}
\end{gather}%
The element $x_{i}$ of the vector $x$ is the probability of playing certain
strategy or the relative frequency of individuals using that strategy. The
fitness function $f_{i}=\sum_{j=1}^{n}a_{ij}x_{j}$ specifies how successful
each subpopulation is, $\left\langle f(x)\right\rangle
=\sum_{k,l=1}^{n}a_{kl}x_{k}x_{l}$ is the average fitness of the population,
and $a_{ij}$ are the elements of the payoff matrix $A$. The replicator
dynamics rewards strategies that outperform the average by increasing their
frequency, and penalizes poorly performing strategies by decreasing their
frequency. The stable fixed points of the replicator dynamics are Nash
equilibria \cite{2}. If a population reaches a state which is a Nash
equilibrium, it will remain there.

\bigskip 

The bonestone of EGT is the concept of evolutionary stable strategy (ESS) 
\cite{4,11} that is a strengthened notion of Nash equilibrium. It satisfies
the following conditions%
\begin{gather}
E(p,p)>E(r,p)\text{,}  \notag \\
\text{If }E(p,p)=E(r,p)\text{ then }E(p,r)>E(r,r)\text{,}  \label{7}
\end{gather}%
where $p$ is the strategy played by the vast majority of the population, and 
$r$ is the strategy of a mutant present in small frequency. Both $p$ and $r$
can be pure or mixed. An ESS is described as a strategy which has the
property that if all the members of a population adopt it, no mutant
strategy could invade the population under the influence of natural
selection. If a few individuals which play a different strategy are
introduced into a population in an ESS, the evolutionary selection process
would eventually eliminate the invaders.

\section{Relationships between Quantum Mechanics \& Game Theory}

A physical or a socioeconomical system (described through quantum mechanics
or game theory) is composed by $n$ members (particles, subsystems, players,
states, etc.). Each member is described by a state or a strategy which has
assigned a determined probability ($x_{i}$ or $\rho _{ij}$). The quantum
mechanical system is described by the density operator $\rho $ whose
elements represent the system average probability of being in a determined
state. In evolutionary game theory the system is defined through a relative
frequencies vector $x$ whose elements can represent the frequency of players
playing a determined strategy. The evolution of the density operator is
described by the von Neumann equation which is a generalization of the Schr%
\"{o}dinger equation. While the evolution of the relative frequencies is
described through the replicator dynamics (\ref{5}).

\bigskip 

It is important to note that the replicator dynamics is a vectorial
differential equation while von Neumann equation can be represented in
matrix form. If we would like to compare both systems the first we would
have to do is to try to compare their evolution equations by trying to find
a matrix representation of the replicator dynamics \cite{12}%
\begin{equation}
\frac{dX}{dt}=G+G^{T}\text{,}  \label{8}
\end{equation}%
where the matrix $X$ has as elements%
\begin{equation}
x_{ij}=\left( x_{i}x_{j}\right) ^{1/2}  \label{9}
\end{equation}%
and%
\begin{eqnarray}
\left( G+G^{T}\right) _{ij} &=&\frac{1}{2}\sum_{k=1}^{n}a_{ik}x_{k}x_{ij} 
\notag \\
&&+\frac{1}{2}\sum_{k=1}^{n}a_{jk}x_{k}x_{ji}  \notag \\
&&-\sum_{k,l=1}^{n}a_{kl}x_{k}x_{l}x_{ij}  \label{10}
\end{eqnarray}%
are the elements of the matrix $\left( G+G^{T}\right) $.

\bigskip 

Although equation (\ref{8}) is the matrix representation of the replicator
dynamics from which we could compare and find a relationship with the von
Neumann equation, we can moreover find a Lax representation of the
replicator dynamics by calling%
\begin{gather}
\left( G_{1}\right) _{ij}=\frac{1}{2}\sum_{k=1}^{n}a_{ik}x_{k}x_{ij}\text{,}
\label{11} \\
\left( G_{2}\right) _{ij}=\frac{1}{2}\sum_{k=1}^{n}a_{jk}x_{k}x_{ji}\text{,}
\label{12} \\
\left( G_{3}\right) _{ij}=\sum_{k,l=1}^{n}a_{kl}x_{k}x_{l}x_{ij}  \label{13}
\end{gather}%
the elements of the matrixes $G_{1}$, $G_{2}$ and $G_{3}$ that compose by
adding the matrix $\left( G+G^{T}\right) $. The matrixes $G_{1},G_{2}$ and $%
G_{3}$ can be also factorized in function of the matrixes $Q$ and $X$%
\begin{gather}
G_{1}=QX\text{,}  \label{14} \\
G_{2}=XQ\text{,}  \label{15} \\
G_{3}=2XQX\text{,}  \label{16}
\end{gather}%
where $Q$ is a diagonal matrix and has as elements $q_{ii}=\frac{1}{2}%
\sum_{k=1}^{n}a_{ik}x_{k}$. By using the fact that $X^{2}=X$ we can write
the equation (\ref{8}) like%
\begin{equation}
\frac{dX}{dt}=QXX+XXQ-2XQX  \label{17}
\end{equation}%
and finally, by grouping into commutators and defining $\Lambda =\left[ Q,X%
\right] $%
\begin{equation}
\frac{dX}{dt}=\left[ \Lambda ,X\right] \text{.}  \label{18}
\end{equation}%
The matrix $\Lambda $ has as elements%
\begin{equation*}
(\Lambda )_{ij}=\frac{1}{2}\left[ \left( \sum_{k=1}^{n}a_{ik}x_{k}\right)
x_{ij}-x_{ji}\left( \sum_{k=1}^{n}a_{jk}x_{k}\right) \right] .
\end{equation*}%
This matrix commutative form of the replicator dynamics (\ref{18}) follows
the same dynamic as the von Neumann equation (\ref{3}) and the properties of
their correspondent elements (matrixes) are similar, being the properties
corresponding to our quantum system more general than the properties of the
classical system.

\bigskip 

The next table shows some specific resemblances between quantum statistical
mechanics and evolutionary game theory \cite{13}.

{\scriptsize Table 1}

\begin{center}
\begin{tabular}{cc}
\hline
{\scriptsize Quantum Statistical Mechanics} & {\scriptsize Evolutionary Game
Theory} \\ \hline
{\scriptsize n system members} & {\scriptsize n population members} \\ 
{\scriptsize Each member in the state }$\left\vert \Psi _{k}\right\rangle $
& {\scriptsize Each member plays strategy }$s_{i}$ \\ 
$\left\vert \Psi _{k}\right\rangle $ {\scriptsize with} $p_{k}\rightarrow $
\ $\rho _{ij}${\scriptsize \ } & $s_{i}${\scriptsize \ }$\ \ \rightarrow $%
{\scriptsize \ }$\ \ x_{i}$ \\ 
$\rho ,$ $\ \ \tsum_{i}\rho _{ii}{\scriptsize =1}$ & ${\scriptsize X,}$%
{\scriptsize \ \ }$\tsum_{i}x_{i}{\scriptsize =1}$ \\ 
${\scriptsize i\hbar }\frac{d\rho }{dt}{\scriptsize =}\left[ \hat{H},\rho %
\right] $ & $\frac{dX}{dt}{\scriptsize =}\left[ \Lambda ,X\right] $ \\ 
${\scriptsize S=-Tr}\left\{ {\scriptsize \rho }\ln {\scriptsize \rho }%
\right\} $ & ${\scriptsize H=-}\tsum\nolimits_{i}{\scriptsize x}_{i}\ln 
{\scriptsize x}_{i}$ \\ 
&  \\ \hline
\end{tabular}
\end{center}

\bigskip 

In table 2 we show the properties of the matrixes $\rho $ and $X$.

{\scriptsize Table 2}

\begin{center}
$%
\begin{tabular}{cc}
\hline
{\scriptsize Density Operator} & {\scriptsize Relative freq. Matrix} \\ 
\hline
$\rho ${\scriptsize \ is Hermitian} & $X${\scriptsize \ is Hermitian} \\ 
${\scriptsize Tr\rho (t)=1}$ & ${\scriptsize TrX=1}$ \\ 
${\scriptsize \rho }^{2}{\scriptsize (t)\leqslant \rho (t)}$ & ${\scriptsize %
X}^{2}{\scriptsize =X}$ \\ 
${\scriptsize Tr\rho }^{2}{\scriptsize (t)\leqslant 1}$ & ${\scriptsize TrX}%
^{2}{\scriptsize (t)=1}$ \\ 
&  \\ \hline
\end{tabular}%
$
\end{center}

Although both systems are different, both are analogous and thus exactly
equivalents.

\section{Quantum Replicator Dynamics \& the Quantization Relationships}

The resemblances between both systems and the similarity in the properties
of their corresponding elements let us to define and propose the next
quantization relationships%
\begin{gather}
x_{i}\rightarrow \sum_{k=1}^{n}\left\langle i\left\vert \Psi _{k}\right.
\right\rangle p_{k}\left\langle \Psi _{k}\left\vert i\right. \right\rangle
=\rho _{ii}\text{,}  \notag \\
(x_{i}x_{j})^{1/2}\rightarrow \sum_{k=1}^{n}\left\langle i\left\vert \Psi
_{k}\right. \right\rangle p_{k}\left\langle \Psi _{k}\left\vert j\right.
\right\rangle =\rho _{ij}\text{.}  \label{19}
\end{gather}%
A population will be represented by a quantum system in which each
subpopulation playing strategy $s_{i}$ will be represented by a pure
ensemble in the state $\left\vert \Psi _{k}(t)\right\rangle $ and with
probability $p_{k}$. The probability $x_{i}$ of playing strategy $s_{i}$ or
the relative frequency of the individuals using strategy $s_{i}$ in that
population will be represented as the probability $\rho _{ii}$ of finding
each pure ensemble in the state $\left\vert i\right\rangle $ \cite{12}.

\bigskip 

Through these quantization relationships the replicator dynamics (in matrix
commutative form) (\ref{18}) takes the form of the equation of evolution of
mixed states (\ref{3}). And also%
\begin{gather}
X\longrightarrow \rho \text{,}  \label{20} \\
\Lambda \longrightarrow -\frac{i}{\hbar }\hat{H}\text{,}  \label{21}
\end{gather}%
where $\hat{H}$ is the Hamiltonian of the physical system.

\bigskip 

The equation of evolution of mixed states from quantum statistical mechanics
(\ref{3}) is the quantum analogue of \ the replicator dynamics in matrix
commutative form (\ref{18}).

\section{Games through Statistical Mechanics \& QIT}

There exists a strong relationship between game theories, statistical
mechanics and information theory. The bonds between these theories are the
density operator and entropy \cite{14,15}. From the density operator we can
construct and understand the statistical behavior about our system by using
the statistical mechanics. Also we can develop the system in function of its
accessible information and analyze it through information theories under a
criterion of maximum or minimum entropy.

\bigskip 

Entropy is the central concept of information theories \cite{16,17}. The
Shannon entropy expresses the average information we expect to gain on
performing a probabilistic experiment of a random variable $A$ which takes
the values $a_{i}$ with the respective probabilities $p_{i}$. It also can be
seen as a measure of uncertainty before we learn the value of $A$. We define
the Shannon entropy of a random variable $A$ by%
\begin{equation}
H(A)\equiv H(p_{1},...,p_{n})\equiv -\sum_{i=1}^{n}p_{i}\log _{2}p_{i}\text{.%
}  \label{22}
\end{equation}%
The entropy of a random variable is completely determined by the
probabilities of the different possible values that the random variable
takes. Due to the fact that $p=(p_{1},...,p_{n})$ is a probability
distribution, it must satisfy $\sum_{i=1}^{n}p_{i}=1$ and $0\leq
p_{1},...,p_{n}\leq 1$. The Shannon entropy of the probability distribution
associated with the source gives the minimal number of bits that are needed
in order to store the information produced by a source, in the sense that
the produced string can later be recovered.

\bigskip 

The von Neumann entropy \cite{16,17} is the quantum analogue of Shannon's
entropy but it appeared 21 years before and generalizes Boltzmann's
expression. Entropy\ in quantum information theory plays prominent roles in
many contexts, e.g., in studies of the classical capacity of a quantum
channel \cite{18,19} and the compressibility of a quantum source \cite{20,21}%
. Quantum information theory appears to be the basis for a proper
understanding of the emerging fields of quantum computation \cite{22,23},
quantum communication \cite{24,25}, and quantum cryptography \cite{26,27}.

\bigskip 

Suppose $A$ and $B$ are two random variables. The joint entropy $H(A,B)$
measures our total uncertainty about the pair $(A,B)$ and it is defined by%
\begin{equation}
H(A,B)\equiv -\sum_{i,j}p_{ij}\log _{2}p_{ij}  \label{23}
\end{equation}%
while%
\begin{gather}
H(A)=-\sum_{i,j}p_{ij}\log _{2}\sum_{j}p_{ij}\text{,}  \label{24} \\
H(B)=-\sum_{i,j}p_{ij}\log _{2}\sum_{i}p_{ij}\text{,}  \label{25}
\end{gather}%
where $p_{ij}$ is the joint probability to find $A$ in state $a_{i}$ and $B$
in state $b_{j}$.

\bigskip 

The conditional entropy $H(A\mid B)$ is a measure of how uncertain we are
about the value of $A$, given that we know the value of $B$. The entropy of $%
A$ conditional on knowing that $B$ takes the value $b_{j}$ is defined by%
\begin{gather}
H(A\mid B)\equiv H(A,B)-H(B)\text{,}  \notag \\
H(A\mid B)\equiv -\sum_{i,j}p_{ij}\log _{2}p_{i\mid j}\text{,}  \label{26}
\end{gather}%
where $p_{i\mid j}=\frac{p_{ij}}{\sum_{i}p_{ij}}$ is the conditional
probability that $A$ is in state $a_{i}$ given that $B$ is in state $b_{j}$.

\bigskip 

The mutual or correlation entropy $H(A:B)$ measures how much information $A$
and $B$ have in common. The mutual or correlation entropy $H(A:B)$ is
defined by%
\begin{gather}
H(A:B)\equiv H(A)+H(B)-H(A,B)\text{,}  \notag \\
H(A:B)\equiv -\sum_{i,j}p_{ij}\log _{2}p_{i:j}\text{,}  \label{27}
\end{gather}%
where $p_{i:j}=\frac{\sum_{i}p_{ij}\sum_{j}p_{ij}}{p_{ij}}$ is the mutual
probability. The mutual or correlation entropy also can be expressed through
the conditional entropy via%
\begin{gather}
H(A:B)=H(A)-H(A\mid B)\text{,}  \label{28} \\
H(A:B)=H(B)-H(B\mid A)\text{.}  \label{29}
\end{gather}%
The joint entropy would equal the sum of each of $A$'s and $B$'s entropies
only in the case that there are no correlations between $A$'s and $B$'s
states. In that case, the mutual entropy or information vanishes and we
could not make any predictions about $A$ just from knowing something about $B
$.

\bigskip 

The relative entropy $H(p\parallel q)$ measures the closeness of two
probability distributions, $p$ and $q$, defined over the same random
variable $A$. We define the relative entropy of $p$ with respect to $q$ by%
\begin{gather}
H(p\parallel q)\equiv \sum_{i}p_{i}\log _{2}p_{i}-\sum_{i}p_{i}\log _{2}q_{i}%
\text{,}  \notag \\
H(p\parallel q)\equiv -H(A)-\sum_{i}p_{i}\log _{2}q_{i}\text{.}  \label{30}
\end{gather}%
The relative entropy is non-negative, $H(p\parallel q)\geq 0$, with equality
if and only if $p=q$. The classical relative entropy of two probability
distributions is related to the probability of distinguishing the two
distributions after a large but finite number of independent samples
(Sanov's theorem) \cite{28}.

\bigskip 

By analogy with the Shannon entropies it is possible to define conditional,
mutual and relative quantum entropies. Quantum entropies also satisfies many
other interesting properties that do not satisfy their classical analogues.
For example, the conditional entropy can be negative and its negativity
always indicates that two systems are entangled and indeed, how negative the
conditional entropy is provides a lower bound on how entangled the two
systems are \cite{29}.

\bigskip 

By other hand, in statistical mechanics entropy can be regarded as a
quantitative measure of disorder. It takes its maximum possible value in a
completely random ensemble in which all quantum mechanical states are
equally likely and is equal to zero in the case of a pure ensemble which has
a maximum amount of order because all members are characterized by the same
quantum mechanical state ket.

\bigskip 

From both possible points of view and analysis (statistical mechanics or
information theories) of the same system its entropy is exactly the same.
Lets consider a system composed by $N$ members, players, strategies, states,
etc. This system is described completely through certain density operator $%
\rho $, its evolution equation (the von Neumann equation) and its entropy.
Classically, the system is described through the matrix of relative
frequencies $X$, the replicator dynamics and the Shannon entropy. For the
quantum case we define the von Neumann entropy as%
\begin{equation}
S=-Tr\left\{ \rho \ln \rho \right\}   \label{31}
\end{equation}%
and for the classical case 
\begin{equation}
H=-\sum_{i=1}x_{ii}\ln x_{ii}  \label{32}
\end{equation}%
which is the Shannon entropy over the relative frequencies vector $x$ (the
diagonal elements of $X$).

\bigskip 

We can describe the evolution of the entropy of our classical system $H(t)$
by supposing that the vector of relative frequencies $x(t)$ evolves in time
following the replicator dynamics \cite{15} 
\begin{equation}
\frac{dH}{dt}=Tr\left\{ U(\tilde{H}-X)\right\} \text{,}  \label{33}
\end{equation}%
where $\tilde{H}$ is a diagonal matrix whose trace is equal to the Shannon
entropy i.e. $H=Tr\tilde{H}$ and $U_{i}=\left[ f_{i}(x)-\left\langle
f(x)\right\rangle \right] $.

\bigskip 

In a far from equilibrium system the von Neumann vary in time until it
reaches its maximum value. When the dynamics is chaotic the variation with
time of the physical entropy goes through three successive, roughly
separated stages \cite{30}. In the first one, $S(t)$ is dependent on the
details of the dynamical system and of the initial distribution, and no
generic statement can be made. In the second stage, $S(t)$ is a linear
increasing function of time ($\frac{dS}{dt}=const.$). In the third stage, $%
S(t)$ tends asymptotically towards the constant value which characterizes
equilibrium ($\frac{dS}{dt}=0$). With the purpose of calculating the time
evolution of entropy we approximate the logarithm of $\rho $ by series $\ln
\rho =(\rho -I)-\frac{1}{2}(\rho -I)^{2}+\frac{1}{3}(\rho -I)^{3}$... \cite%
{15} 
\begin{eqnarray}
\frac{dS(t)}{dt} &=&\frac{11}{6}\tsum\limits_{i}\frac{d\rho _{ii}}{dt} 
\notag \\
&&-6\tsum\limits_{i,j}\rho _{ij}\frac{d\rho _{ji}}{dt}  \notag \\
&&+\frac{9}{2}\tsum\limits_{i,j,k}\rho _{ij}\rho _{jk}\frac{d\rho _{ki}}{dt}
\notag \\
&&-\frac{4}{3}\tsum\limits_{i,j,k,l}\rho _{ij}\rho _{jk}\rho _{kl}\frac{%
d\rho _{li}}{dt}+\zeta \text{.}  \label{34}
\end{eqnarray}

\bigskip 

In general entropy can be maximized subject to different constrains. In each
case the result is the condition the system must follow to maximize its
entropy. Generally, this condition is a probability distribution function.
We can obtain the density operator from the study of an ensemble in thermal
equilibrium. Nature tends to maximize entropy subject to the constraint that
the ensemble average of the Hamiltonian has a certain prescribed value. We
will maximize $S$ by requiring that%
\begin{equation}
\delta S=-\sum_{i}\delta \rho _{ii}(\ln \rho _{ii}+1)=0  \label{35}
\end{equation}%
subject to the constrains $\delta Tr\left( \rho \right) =0$ and $\delta
\left\langle E\right\rangle =0$. By using Lagrange multipliers%
\begin{equation}
\sum_{i}\delta \rho _{ii}(\ln \rho _{ii}+\beta E_{i}+\gamma +1)=0  \label{36}
\end{equation}%
and the normalization condition $Tr(\rho )=1$ we find that%
\begin{equation}
\rho _{ii}=\frac{e^{-\beta E_{i}}}{\sum_{k}e^{-\beta E_{k}}}  \label{37}
\end{equation}%
which is the condition that the density operator and its elements must
satisfy to our system tends to maximize its entropy $S$. If we maximize $S$
without the internal energy constrain $\delta \left\langle E\right\rangle =0$
we obtain%
\begin{equation}
\rho _{ii}=\frac{1}{N}  \label{38}
\end{equation}%
which is the $\beta \rightarrow 0$\ limit (\textquotedblleft high -
temperature limit\textquotedblright ) in equation (\ref{37}) in where a
canonical ensemble becomes a completely random ensemble in which all energy
eigenstates are equally populated. In the opposite low - temperature limit $%
\beta \rightarrow \infty $ tell us that a canonical ensemble becomes a pure
ensemble where only the ground state is populated. The parameter $\beta $ is
related to the \textquotedblleft temperature\textquotedblright\ $\tau $ as
follows%
\begin{equation}
\beta =\frac{1}{\tau }\text{.}  \label{39}
\end{equation}%
By replacing $\rho _{ii}$ obtained in the equation (\ref{37}) in the von
Neumann entropy we can rewrite it in function of the partition function $%
Z=\sum_{k}e^{-\beta E_{k}}$, $\beta $ and $\left\langle E\right\rangle $
through the next equation%
\begin{equation}
S=\ln Z+\beta \left\langle E\right\rangle \text{.}  \label{40}
\end{equation}%
From the partition function we can know some parameters that define the
system like%
\begin{gather}
\left\langle E\right\rangle =-\frac{1}{Z}\frac{\partial Z}{\partial \beta }=-%
\frac{\partial \ln Z}{\partial \beta }\text{,}  \label{41} \\
\left\langle \Delta E^{2}\right\rangle =-\frac{\partial \left\langle
E\right\rangle }{\partial \beta }=-\frac{1}{\beta }\frac{\partial S}{%
\partial \beta }\text{.}  \label{42}
\end{gather}%
We can also analyze the variation of entropy with respect to the average
energy of the system%
\begin{gather}
\frac{\partial S}{\partial \left\langle E\right\rangle }=\frac{1}{\tau }%
\text{,}  \label{43} \\
\frac{\partial ^{2}S}{\partial \left\langle E\right\rangle ^{2}}=-\frac{1}{%
\tau ^{2}}\frac{\partial \tau }{\partial \left\langle E\right\rangle }
\label{44}
\end{gather}%
and with respect to the parameter $\beta $%
\begin{gather}
\frac{\partial S}{\partial \beta }=-\beta \left\langle \Delta
E^{2}\right\rangle \text{,}  \label{45} \\
\frac{\partial ^{2}S}{\partial \beta ^{2}}=\frac{\partial \left\langle
E\right\rangle }{\partial \beta }+\beta \frac{\partial ^{2}\left\langle
E\right\rangle }{\partial \beta ^{2}}\text{.}  \label{46}
\end{gather}

\section{From Classical to Quantum}

The resemblances between both systems (described through quantum mechanics
and EGT) apparently different but analogous and thus exactly equivalents and
the similarity in the properties of their corresponding elements let us to
define and propose the quantization relationships like in section 5.

\bigskip 

It is important to note that equation (\ref{18}) is nonlinear while its
quantum analogue is linear. This means that the quantization eliminates the
nonlinearities. Also through this quantization the classical system that
were described through a diagonal matrix $X$ can be now described through a
density operator which not neccesarily must describe a pure state, i.e. its
non diagonal elements can be different from zero representing a mixed state
due to the coherence between quantum states that were not present through a
classical analysis.

\bigskip 

Through the relationships between both systems we could describe classical,
evolutionary, quantum and also the biological systems that were described
before through evolutionary dynamics with the replicator dynamics. We could
explain through quantum mechanics biological and economical processes being
a much more general theory that we had thought. It could even encloses
theories like games and evolutionary dynamics.

\bigskip 

Problems in economy and finance have attracted the interest of statistical
physicists. Kobelev et al \cite{31} used methods of statistical physics of
open systems for describing the time dependence of economic characteristics
(income, profit, cost, supply, currency, etc.) and their correlations with
each other. Antoniou et al \cite{32} introduced a new approach for the
presentation of economic systems with a small number of components as a
statistical system described by density functions and entropy. This analysis
is based on a Lorenz diagram and its interpolation by a continuos function.
Conservation of entropy in time may indicate the absence of macroscopic
changes in redistribution of resources. Assuming the absence of
macro-changes in economic systems and in related additional expenses of
resources, we may consider the entropy as an indicator of efficiency of the
resources distribution. Statistical physicists are also extremely interested
in economic fluctuations \cite{33} in order to help our world financial
system avoid \textquotedblleft economic earthquakes\textquotedblright . Also
it is suggested that in the field of turbulence, we may find some crossover
with certain aspects of financial markets. Statistical mechanics and
economics study big ensembles: collections of atoms or economic agents,
respectively. The fundamental law of equilibrium statistical mechanics is
the Boltzmann-Gibbs law, which states that the probability distribution of
energy $E$ is $P(E)=Ce^{-E/T}$, where $T$ is the temperature, and $C$ is a
normalizing constant. The main ingredient that is essential for the
derivation of the Boltzmann-Gibbs law is the conservation of energy. Thus,
one may generalize that any conserved quantity in a big statistical system
should have an exponential probability distribution in equilibrium \cite{34}%
. In a closed economic system, money is conserved. Thus, by analogy with
energy, the equilibrium probability distribution of money must follow the
exponential Boltzmann-Gibbs law characterized by an effective temperature
equal to the average amount of money per economic agent. Dr\u{a}gulescu and
Yakovenko demonstrated how the Boltzmann-Gibbs distribution emerges in
computer simulations of economic models. They considered a thermal machine,
in which the difference of temperature allows one to extract a monetary
profit. They also discussed the role of debt, and models with broken
time-reversal symmetry for which the Boltzmann-Gibbs law does not hold.
Recently the insurance market, which is one of the important branches of
economy, have attracted the attention of physicists \cite{35}. The maximum
entropy principle is used for pricing the insurance. Darooneh obtained the
price density based on this principle, applied it to multi agents model of
insurance market and derived the utility function. The main assumption in
his work is the correspondence between the concept of the equilibrium in
physics and economics. He proved that economic equilibrium can be viewed as
an asymptotic approximation to physical equilibrium and some difficulties
with mechanical picture of the equilibrium may be improved by considering
the statistical description of it. Tops{\scriptsize \O }e \cite{36} also has
suggested that thermodynamical equilibrium equals game theoretical
equilibrium. Quantum games have proposed a new point of view for the
solution of the classical problems and dilemmas in game theory. Quantum
games are more efficient than classical games and provide a saturated upper
bound for this efficiency \cite{37,38,39,40,41,42}.

\bigskip 

Nature may be playing quantum survival games at the molecular level \cite%
{43,44}. It could lead us to describe many of the life processes through
quantum mechanics like Gogonea and Merz \cite{45} who indicated that games
are being played at the quantum mechanical level in protein folding.
Gafiychuk and Prykarpatsky \cite{46} applied the replicator equations
written in the form of nonlinear von Neumann equations to the study of the
general properties of the quasispecies dynamical system from the standpoint
of its evolution and stability. They developed a mathematical model of a
naturally fitted coevolving ecosystem and a theoretical study a
self-organization problem of an ensemble of interacting species. \ The
genetic code is the relationship between the sequence of the bases in the
DNA and the sequence of amino acids in proteins. Recent work \cite{47} about
evolvability of the genetic code suggests that the code is shaped by natural
selection. DNA is a nonlinear dynamical system and its evolution is a
sequence of chemical reactions. An abstract DNA-type system is defined by a
set of nonlinear kinetic equations with polynomial nonlinearities that admit
soliton solutions associated with helical geometry. Aerts and Czachor \cite%
{48} shown that the set of these equations allows for two different Lax
representations: They can be written as von Neumann type nonlinear systems
and they can be regarded as a compatibility condition for a
Darboux-covariant Lax pair. Organisms whose DNA evolves in a chaotic way
would be eliminated by natural selection.\ They also explained why
non-Kolmogorovian probability models occurring in soliton kinetics are
naturally associated with chemical reactions. Patel \cite{49,50} suggested
quantum dynamics played a role in the DNA replication and the optimization
criteria involved in genetic information processing. He considers the
criteria involved as a task similar to an unsorted assembly operation where
the Grover's database search algorithm fruitfully applies; given the
different optimal solutions for classical and quantum dynamics. Turner and
Chao \cite{51} studied the evolution of competitive interactions among
viruses in an RNA phage, and found that the fitness of the phage generates a
payoff matrix conforming to the two-person prisoner's dilemma game.
Bacterial infections by viruses have been presented as classical game-like
situations where nature prefers the dominant strategies. Azhar Iqbal \cite%
{42} showed results in which quantum mechanics has strong and important
roles in selection of stable solutions in a system of interacting entities.
These entities can do quantum actions on quantum states. It may simply
consists of a collection of molecules and the stability of solutions or
equilibria can be affected by quantum interactions which provides a new
approach towards theories of rise of complexity in groups of quantum
interacting entities. Neuroeconomics \cite{52,53} may provide an alternative
to the classical Cartesian model of the brain and behavior \cite{54} through
a rich dialogue between theoretical neurobiology and quantum logic \cite%
{55,56}.

\bigskip 

The results shown in this study on the relationships between quantum
mechanics and game theories are a reason of the applicability of physics in
economics and biology. Both systems described through two apparently
different theories are analogous and thus exactly equivalents. So, we can
take some concepts and definitions from quantum mechanics and physics for
the best understanding of the behavior of economics and biology. Also, we
could maybe understand nature like a game in where its players compete for a
common welfare and the equilibrium of the system that they are members.

\section{On a Quantum Understanding of Classical Systems}

If our systems are analogous and thus exactly equivalents, our physical
equilibrium (maximum entropy) should be also exactly equivalent to our
socieconomical equilibrium. If in an isolated system each of its accessible
states do not have the same probability, the system is not in equilibrium.
The system will vary and will evolution in time until it reaches the
equilibrium state in where the probability of finding the system in each of
the accessible states is the same. The system will find its more probable
configuration in which the number of accessible states is maximum and
equally probable. The whole system will vary and rearrange its state and the
states of its ensembles with the purpose of maximize its entropy and reach
its maximum entropy state. We could say that the purpose and maximum payoff
of a physical system is its maximum entropy state. The system and its
members will vary and rearrange themselves to reach the best possible state
for each of them which is also the best possible state for the whole system.

\bigskip 

This can be seen like a microscopical cooperation between quantum objects to
improve their states with the purpose of reaching or maintaining the
equilibrium of the system. All the members of our quantum system will play a
game in which its maximum payoff is the equilibrium of the system. The
members of the system act as a whole besides individuals like they obey a
rule in where they prefer the welfare of the collective over the welfare of
the individual. This equilibrium is represented in the maximum system
entropy in where the system resources\ are fairly distributed over its
members. A system is stable only if it maximizes the welfare of the
collective above the welfare of the individual. If it is maximized the
welfare of the individual above the welfare of the collective the system
gets unstable and eventually it collapses (Collective Welfare Principle\ 
\cite{12,13,15}).

\bigskip 

Fundamentally, we could distinguish three states in every system: minimum
entropy, maximum entropy, and when the system is tending to whatever of
these two states. The natural trend of a physical system is to the maximum
entropy state. The minimum entropy state is a characteristic of a
manipulated\ system i.e. externally controlled or imposed. A system can be
internally or externally manipulated or controlled with the purpose of guide
it to a state of maximum or minimum entropy depending of the ambitions of
the members that compose it or the people\ who control it.

\bigskip 

There exists tacit rules inside a system. These rules do not need to be
specified or clarified and search the system equilibrium under the
collective welfare principle. The other prohibitive\ and repressive\ rules
are imposed over the system when one or many of its members violate the
collective welfare principle and search to maximize its individual welfare
at the expense of the group. Then it is necessary to establish regulations
on the system to try to reestablish the broken natural order.

\section{Conclusions}

The relationships between game theory and quantum mechanics let us propose
certain quantization relationships through which we could describe and
understand not only classical and evolutionary systems but also the
biological systems that were described before through the replicator
dynamics. Quantum mechanics could be used to explain more correctly
biological and economical processes and even encloses theories like games
and evolutionary dynamics.

\bigskip 

The quantum analogues of the relative frequencies matrix, the replicator
dynamics and the Shannon entropy are the density operator, the von Neumann
equation and the von Neumann entropy. Every game (classical, evolutionary or
quantum) can be described quantically through these three elements.

\bigskip 

The bonds between game theories, statistical mechanics and information
theory are the density operator and entropy. From the density operator we
can construct and obtain all the mechanical statistical information about
our system. Also we can develop the system in function of its information
and analyze it through information theories under a criterion of maximum or
minimum entropy.

\bigskip 

Although both systems analyzed are described through two apparently
different theories (quantum mechanics and game theory) both are analogous
and thus exactly equivalents. So, we can take some concepts and definitions
from quantum mechanics and physics for the best understanding of the
behavior of economics and biology. Also, we could maybe understand nature
like a game in where its players compete for a common welfare and the
equilibrium of the system that they are members.

\bigskip 

We could say that the purpose and maximum payoff of a system is its maximum
entropy state. The system and its members will vary and rearrange themselves
to reach the best possible state for each of them which is also the best
possible state for the whole system. This can be seen like a microscopical
cooperation between quantum objects to improve their states with the purpose
of reaching or maintaining the equilibrium of the system. All the members of
our system will play a game in which its maximum payoff is the equilibrium
of the system. The members of the system act as a whole besides individuals
like they obey a rule in where they prefer to work for the welfare of the
collective besides the individual welfare.

\end{document}